\documentclass[lettersize,journal]{IEEEtran}
\usepackage{amsmath,amsfonts}
\usepackage{algorithmic}
\usepackage{algorithm}
\usepackage{array}
\usepackage[caption=false,font=normalsize,labelfont=sf,textfont=sf]{subfig}
\usepackage{textcomp}
\usepackage{stfloats}
\usepackage{url}
\usepackage{verbatim}
\usepackage{graphicx}
\usepackage{cite}
\usepackage{hyperref}
\usepackage[all]{hypcap}
\usepackage{color}
\usepackage{booktabs}
\begin{document}

\title{Hybrid HMM Decoder For Convolutional Codes By Joint Trellis-Like Structure and Channel Prior}

\author{Haoyu Li, Xuan Wang, Tong Liu, Dingyi Fang,~\IEEEmembership{Member,~IEEE}, Baoying Liu%
\thanks{Haoyu Li, Xuan Wang, Tong Liu, Dingyi Fang, Baoying Liu are with School of Information Science and Technology, Northwest University, Xi'an, 710127, China 
(email: lihaoyu@stumail.nwu.edu.cn; xwang@stumail.nwu.edu.cn; liutong98@stumail.nwu.edu.cn; dyf@nwu. edu.cn; paola.liu@nwu.edu.cn).}
\thanks{Code is available at \url{https://github.com/HaoyyLi/HMM-decoder}
}
}

\markboth{IEEE TRANSACTIONS ON COGNITIVE COMMUNICATIONS AND NETWORKING}%
{Shell \MakeLowercase{\textit{et al.}}: Hybrid HMM Decoder For Convolutional Codes By Joint Trellis-Like Structure and Channel Prior}



\maketitle

\begin{abstract}
  The anti-interference capability of wireless links is a physical layer problem for edge computing.
  Although convolutional codes have inherent error correction potential due to the redundancy introduced in the data, the performance of the convolutional code is drastically degraded due to multipath effects on the channel.
  In this paper, we propose the use of a Hidden Markov Model (HMM) for the reconstruction of convolutional codes and decoding by the Viterbi algorithm.
  Furthermore, to implement soft-decision decoding, the observation of HMM is replaced by Gaussian mixture models (GMM). 
  Our method provides superior error correction potential than the standard method because the model parameters contain channel state information (CSI).
  We evaluated the performance of the method compared to standard Viterbi decoding by numerical simulation. In the multipath channel, the hybrid HMM decoder can achieve a performance gain of 4.7 dB and 2 dB when using hard-decision and soft-decision decoding, respectively.
  The HMM decoder also achieves significant performance gains for the RSC code, suggesting that the method could be extended to turbo codes.
\end{abstract}


\section{INTRODUCTION}
\label{INTRODUCTION}
Edge computing is a hot research topic that can compensate for the shortcomings of cloud computing and meet the computing needs of densely distributed IoT devices\cite{mach2017mobile,ren2018edge,premsankar2018edge,mao2017survey,shi2016promise}.
Unlike cloud computing, which uses stable wired connections, edge devices wirelessly transmit data to neighboring devices to perform computations. Therefore, the reliability of wireless communication has become a critical issue in edge computing\cite{mao2017survey}. 

The wireless signal is usually affected by multipath channels due to the environment that causes the symbols of the data stream to fuse, i.e. intersymbol interference (ISI).
Furthermore, as the communication distance increases, the signal-to-noise ratio (SNR) degrades rapidly because the wireless signal is attenuated in the air, increasing the bit error rate (BER).
To make matters worse, signal distortions caused by the nonlinear effects of electronic components increase detection errors.

To resist interference, channel coding is implemented that provides an error correction mechanism to reduce bit errors. 
Considering the limited computational resources of edge devices, convolutional codes are most suitable for edge computing applications because of the simple structure and excellent error correction capabilities.
The remarkable performance of implementing convolutional codes on low-power communication devices was demonstrated by\cite{bharadia2015backfi}.

Convolutional coding is a type of memory coding in which a data stream is converted into an encoded stream by a mathematical process.
In Gaussian channels, errors are randomly dispersed, and the Viterbi decoder\cite{viterbi1967error} can recover data stream based on the context of the encoded stream.
However, the ISI caused by the multipath effect may result in continuous bit errors rather than a Gaussian distribution as in Gaussian channels. To maximize the coding gain, the received signal should be equalized based on the channel state information (CSI). Unfortunately, CSI cannot be calculated accurately. This will reduce the performance of the convolutional code.

Researchers have made tremendous efforts to improve the performance of convolutional codes in multipath channels. \cite{hagenauer1989viterbi} proposed the soft-input and soft-output Viterbi algorithm (SOVA) based on the conventional Viterbi decoder. The conventional Viterbi decoder can only decode the binary stream based on the demodulator decision. In contrast, the SOVA is based on the computation of the log-likelihood function (LLF) and soft estimation of the data stream. In this way, prior probabilities can be included and significant coding gains can be achieved. However, SOVA optimizes the performance of convolutional codes by improving the decoding algorithm rather than the model design, and the channel prior is still ignored. %

The Deep Learning-based convolutional decoder uses neural networks to handle the decoding task \cite{aghamalek2013improved,berber2004soft,hamalainen1999recurrent,hamalainen1999convolutional,hueske2007improving,inbook,ou2020neural,kim2018communication}. Since the end-to-end network structure contains the channel's prior information, the neural network decoder shows great decoding performance. However, due to the huge number of the neural network parameters, it takes a long time to decode the data stream. Without GPU, it would take several minutes even for a small test set. Therefore, decoding methods based on deep learning still faces challenges on edge devices that need to be solved.

The purpose of this paper is to develop a method for improving the error correction potential of convolutional codes without increasing the complexity of the decoding algorithm. After analyzing the trellis-like structure of convolutional codes and the Viterbi decoding algorithm, we find that the state transition relation of the convolutional code satisfies the homogeneous Markov property since the data stream is a random process. Thus, the convolutional code can be established as a hidden Markov model (HMM). Based on this intuition, we can optimize the performance of convolutional codes by reconstructing the HMM model.

\begin{figure*}[!t]
  \centering
  \subfloat[]{\includegraphics[width=5.25in]{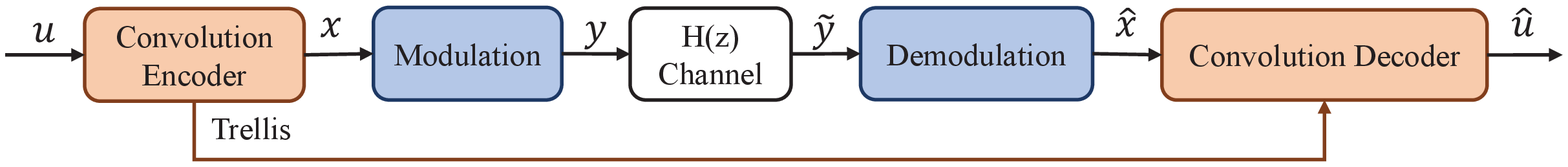}%
  \label{fig1.a}}
  \hfil
  \subfloat[]{\includegraphics[width=5.25in]{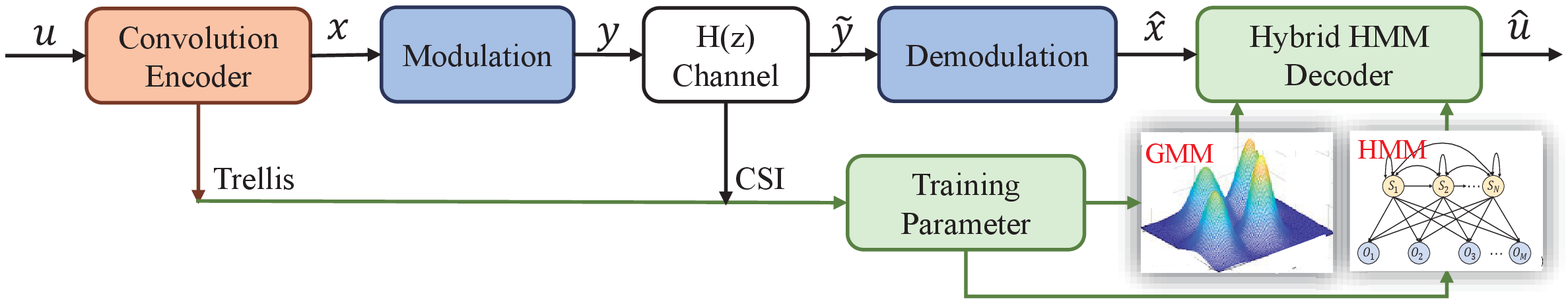}%
  \label{fig1.b}}
  \caption{System framework of (a) conventional Viterbi decoder and (b) hybrid HMM decoder.}
  \label{fig1}
\end{figure*}

A major challenge that has not yet been addressed is that it is difficult to compute perfect CSI. We derive the channel prior using a probability matrix in the model, which circumvents the problem of imperfection CSI. Specifically, the mathematical logic of the convolutional code is a one-to-one mapping between the state of the register and the output of the encoder, described by a Markov process. However, the mapping relationship would be disturbed by multipath channels. In this case, the Markov process is modeled as an HMM and the channel prior is quantified by the HMM observation matrix. Compared with the conventional method shown in \hyperref[fig1.a]{Fig. 1(a)}, our model parameterizes the channel prior information, as shown in \hyperref[fig1.a]{Fig. 1(b)}.

The SOVA decoding approach in \cite{hagenauer1989viterbi} provides superior error correction potential because soft estimation provides the prior probabilities information. However, the discrete HMM, a discrete variables model, cannot recover the data stream based on the soft input sequence. To benefit from the advantages of soft-decision decoding, the likelihood of the demodulator output is estimated by a Gaussian mixture distribution and fitted into a Gaussian mixture model (GMM) as the observation of the hybrid HMM decoder. Unlike the HMM, GMM-HMM is a continuous variable model, so the model can accept continuous observation sequences for decoding. Thus, the GMM-based hybrid HMM model can perform soft-decision decoding.

Numerical simulation results demonstrate that the hybrid HMM decoder is superior to the standard Viterbi decoding algorithm by 4.7 dB and 2 dB performance gains in the hard decision and soft-decision decoding, respectively. Furthermore, unlike neural network decoders, HMM does not require large-scale parameters and the training procedure generally takes less than a minute. Our approach also has high decoding performance on Recursive System Convolution (RSC) codes, indicating that it can also decode Turbo codes.

Our contributions are summarized as follows:

1. We propose a novel HMM decoder that can be trained to parameterize the trellis-like structure of the convolutional code and channel prior information. The error correction performance under multipath channels is superior to the conventional Viterbi decoder.

2. To take advantage of soft-decision and achieve a significant performance improvement, the hybrid HMM based on GMM, a continuous variable model, is developed. The observation in HMM was replaced by GMM, which was trained to fit the distribution of demodulator likelihood to be compatible with soft-decision decoding.

3. The results of the numerical analysis show that the hybrid HMM decoder outperforms the standard Viterbi decoder by 4.7 dB and 2 dB for hard and soft decisions in multipath channels, respectively. We evaluate the effects of different code rates, constraint lengths, and encoding methods on decoding performance.

4. We note that the hybrid HMM decoder offers significant performance gains, suggesting that our technology may be extended to turbo codes in the future.

The remainder of this paper is organized as follows. In Section \ref{RELATED WORK}, we review the related work. Section \ref{BACKGROUND} shows the background related to this work. The framework and process of the system are introduced in Section \ref{SYSTEM OVERVIEW}. Sections \ref{DECODING CONVOLUTIONAL CODES WITH HMM} and \ref{DECODING CONVOLUTIONAL CODES WITH GMMHMM} analyze the details of the method, including HMM 
decoder and GMM-HMM for soft-decision decoding. Section \ref{EXPERIMENTAL RESULT} shows experimental results. The limitations of HMM decoders and future research topics are discussed in Section \ref{DISCUSSION AND LIMITATIONS}. Section \ref{CONCLUSION AND FUTURE WORK} presents the summary and prospects of this work.

\section{RELATED WORK}
\label{RELATED WORK}
\subsection{Machine Learning in Wireless Communications}
\label{Machine Learning in Communications}
In recent years, machine learning, especially Deep Learning based on neural networks, has developed rapidly. Machine learning has been applied to all layers of communication systems to improve performance as it has excellent results in fitting and classification tasks. \cite{wang2019modulation,xiong2019robust,chen2020signet,wang2020lightamc,doan2020learning,qu2019radar,ramjee2019fast,li2018generative} have demonstrated the potential of machine learning methods to recognize modulation methods of communication systems. 
In addition, researchers have developed a machine learning-based approach to symbol detection and demodulation of the wireless signal\cite{qu2020mud,gorday2020gfsk,kozlenko2020software,zhang2020intelligent,shlezinger2019viterbinet}. 
Other popular research areas for machine learning applications in communication systems include signal enhancement\cite{zhou2020wireless} and end-to-end communication\cite{erpek2020deep,kulin2018end}.

\subsection{Channel Decoder based on Machine Learning}
\label{Channel Decoder based on Machine Learning}
The neural network decoder achieves higher coding gain because the regular mapping correlations and trellis-like structure of channel coding can be fitted as nonlinear functions. Researchers obtain a lower BER, by constructing a neural network structure based on a graph of linear codes in \cite{nachmani2018deep,nachmani2016learning}. Complex codes that have better error correction or even reach the Shannon limit can also be decoded by neural networks.
\cite{ye2017initial,xu2018joint} use an end-to-end neural network to simultaneously equalize and decode. TurboAE, an end-to-end Turbo decoder based on the autoencoder neural network, was proposed by \cite{jiang2019turbo} to increase the reliability of the Turbo decoder without sacrificing the coding gain. \cite{liang2018iterative} combined belief-propagation (BP) with CNN and developed a decoder that can be iteratively optimized for low-density parity-check (LDPC) codes. Specifically, BP is used to estimate the data stream, and CNN is used to eliminate estimation errors.

\subsection{Applications of HMM}
\label{Applications of HMM}
The Hidden Markov Model, a statistical model, was first proposed by Baum L.E. in the 1960s. The HMM has been increasingly popular in natural language processing \cite{tanaka2019joint,palaz2019end,sun2020improving,novoa2018uncertainty,satori2017voice}. Furthermore, the HMM is also employed in a variety of other fascinating fields. In \cite{samanta2018hmm}, HMM is used to recognize handwritten symbols online. In the application of physiological signal analysis, HMM can recognize sleep stages based on electroencephalogram (EEG)\cite{jiang2019robust,ghimatgar2020neonatal}. \cite{guo2017eeg} has proposed a novel approach using an SVM-HMM to recognize human emotions based on electroencephalogram (EEG) signals.

\section{BACKGROUND}
\label{BACKGROUND}
\subsection{Convolutional Codes}
\label{Convolutional Codes}
\begin{figure*}[!t]
  \centering
  \subfloat[]{\includegraphics[width=4.1in]{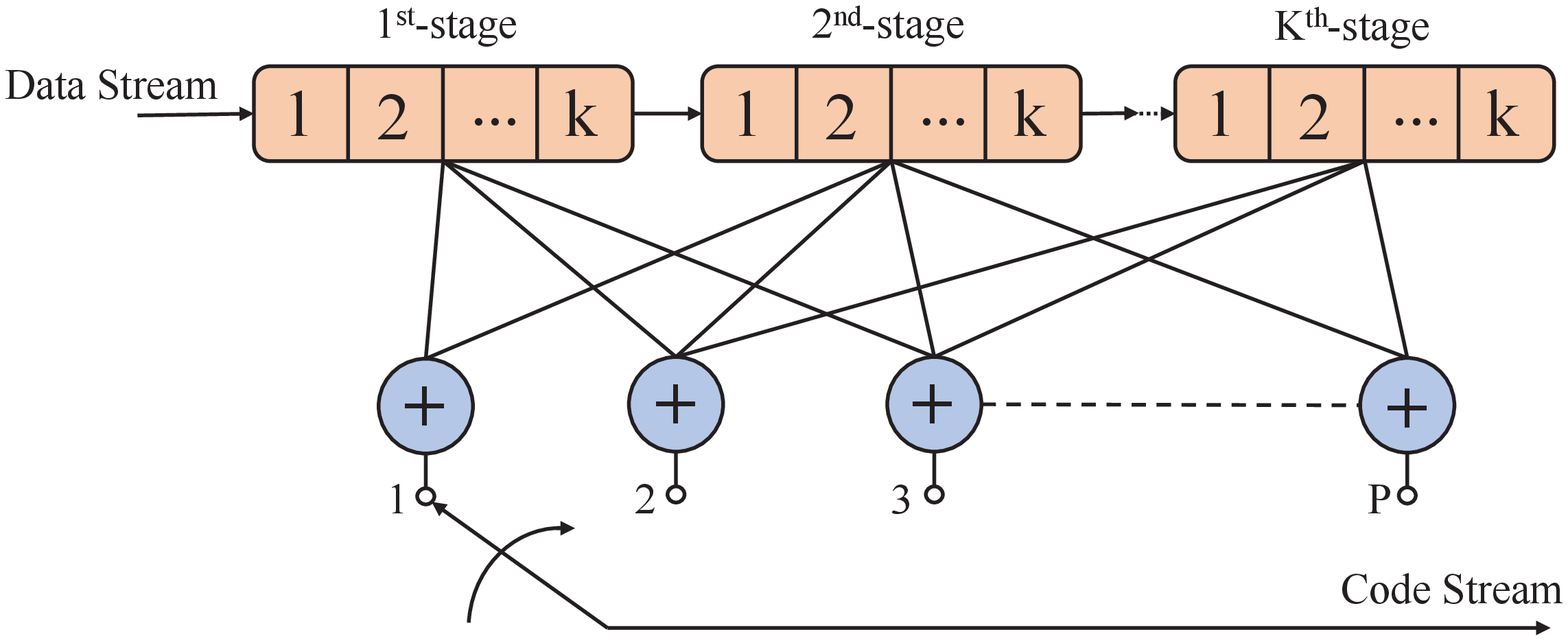}
  \label{fig2.a}}
  \subfloat[]{\includegraphics[width=2.2in]{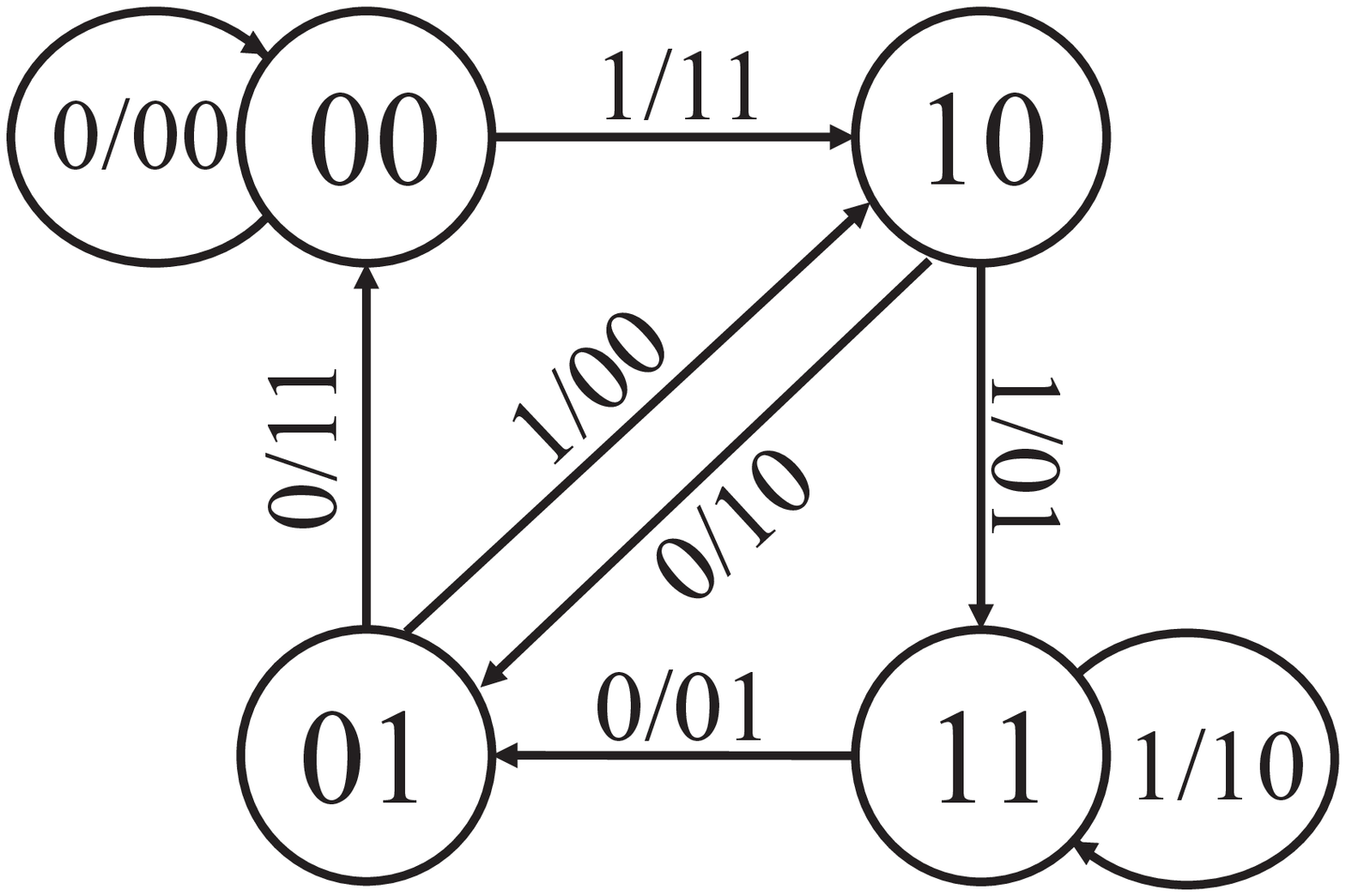}
  \label{fig2.b}}
  \caption{(a) Convolutional code composed of $K$-stage shift registers and $P$ generator polynomial functions; (b) is the state diagram of (2,1,3) convolutional codes}
  \label{fig2}
  \end{figure*}

Convolutional code is a type of error correction code that generates binary symbols by slidingly applying a generator polynomial function to the data stream. Sliding means that the generator polynomial function is convolved with the data stream. \hyperref[fig2.a]{Fig. 2(a)} describes a convolutional encoder $(P,k,K)$ consists of $K$-stage shift registers, which have $k$ bits per stage, and $P$ linear $mod-2$ adders as the generator polynomial function that generates coding streams. As the shift register slides over the data stream with $k$ bits each time, $P$ generator polynomial functions output $P$ bits of sequence. Therefore, the coding rate of the convolutional encoder is calculated as $k/P$. In this article, we usually consider convolutional coding with $k=1$, which means that our convolutional encoder moves one bit each time, and the coding rate is $1/P$. $K$ is the constraint length of the convolutional encoder. Under the condition of $k=1$, the coding result of each bit is jointly determined by the current bit and the previous $K-1$ bits. Thus, the convolutional code is a kind of memory code. The generator polynomial function, expressed as a binary vector $g=\{ 0,1 \}^K$, determines the mapping rules between the input and output of the convolutional encoder. The output bit of function is the $mod-2$ operation with the register value in the shift register: 
\begin{equation}
  y_t = G\cdot x_t\;mod\;2
\end{equation}
where $x_t$ is the state of registers at time $t$. $G$ is a $P\times K$ matrix signifying function generators. $P$ is the number of the generator polynomial function, and $K$ is the dimension of the vector $g$. $y_t$ is the output of the convolutional encoder at time $t$, which is a $P$-dimension vector.

The convolutional code is represented by the state diagram in \hyperref[fig2.a]{Fig. 2(b)}, which shows the state diagram of the $(2,1,3)$ convolutional code with the generator polynomial as $[7,5]$. The value of the shift register is expressed as a state, and the output of the generator polynomial function is produced by the state. This is a Hidden Markov Model which was introduced in Section \ref{Hidden Markov Model}. However, the convolutional code only considers the mathematical connection in the model, and the influence of the channel has not been quantified. 

\subsection{Hidden Markov Model}
\label{Hidden Markov Model}
The Hidden Markov Model(HMM) is a generative statistical learning model that describes the process of generating a sequence of states from a Markov chain and then generating an observation sequence from the sequence of states. The HMM is a model with hidden variables because the observation samples can be counted but the sequence of the state is usually unknown, that is, in a hidden form.

Assume that $I=(i_1,i_2,\cdots,i_T)$ is a sequence of states generated by a Markov chain and that the elements of the sequence $I$ come from the set $Q=\{ q_1,q_2,\cdots,q_N \}$, which is the set of all possible states. The transition relation between the states is expressed by a matrix:
\begin{equation}
  A=\begin{pmatrix}
    a_{11} & a_{12} & \cdots & a_{1N}\\
    a_{21} & a_{22} & \cdots & a_{2N}\\
    \vdots  & \vdots & \ddots & \vdots\\
    a_{N1} & a_{N2} & \cdots & a_{NN}\\
    \end{pmatrix}
\end{equation}
where $a_{ij}$ denotes the probability of transition from state $q_i$ to state $q_j$. The initial state vector is expressed as $\pi=(\pi_1,\pi_2,\cdots,\pi_N)$. The observation sequence $O=(o_1,o_2,\cdots,o_T)$ is generated by the state sequence $I$, and the elements in $O$ come from the observation set $V=\{ v_1,v_2,\cdots,v_M \}$. If each state generates a unique observation outcome, that is, the state and the observation are one-to-one, then we obtain a Markov process. However, there is no one-to-one mapping between the state and the observation but a probability distribution. The matrix $B$ is used to describe this generative relationship:
\begin{equation}
  B=\begin{pmatrix}
    b_{11} & b_{12} & \cdots & b_{1M}\\
    b_{21} & b_{22} & \cdots & b_{2M}\\
    \vdots  & \vdots & \ddots & \vdots\\
    b_{N1} & b_{N2} & \cdots & b_{NM}\\
    \end{pmatrix}
\end{equation}
where $b_{ij}$ is the probability of obtaining the observation $v_j$ from the state $q_i$. The formal definition of an HMM is as follows:
\begin{equation}
  \lambda=(A,B,\pi)
\end{equation}

The hidden Markov model has two basic assumptions:

\textbf{Assumption 1:} Homogeneous Markov Property

Hidden Markov chains obey homogeneous Markov properties. Specifically, the current state of the Markov chain depends only on the previous state and is independent of others.
\begin{equation}
  P(i_t|i_{t-1},o_{t-1},\cdots,i_1,o_1) = P(i_t|i_{t-1}),\quad t=1,2,\cdots,T
\end{equation}

\textbf{Assumption 2:} Observational Independence

The output observation generated by the hidden Markov model at time $t$ depends only on the current state and has nothing to do with other states and observations. 
\begin{equation}
  P(o_t|i_t,i_{t-1},o_{t-1},\cdots,i_1,o_1) = P(o_t|i_t),\quad t=1,2,\cdots,T
\end{equation}

In Section \ref{DECODING CONVOLUTIONAL CODES WITH HMM}, we model the convolutional encoding process as a Hidden Markov Model and use the training algorithm to quantify the model parameters $\lambda=(A, B, \pi)$. Although both the convolutional code and the HMM model are decoded using the Viterbi algorithm\cite{forney1973viterbi}, the HMM with parameterized channel priors shows better decoding performance.

\section{SYSTEM OVERVIEW}
\label{SYSTEM OVERVIEW}
\hyperref[fig1.a]{Fig. 1(a)} shows the basic components of a communication system, including transmitters, receivers, and channels. The data stream $u$ is mapped into the complex domain $y$ after being encoded and modulated. Assuming a channel including both line-of-sight (LOS) and non-line-of-sight (N-LOS) paths, the channel impulse response can be expressed as:
{
    \begin{equation}
      h=(h_0, h_1,\cdots,h_{n-1})
    \end{equation}
}
where $h$ is a sequence, $n$ is the number of paths that can reach the receiver, $h_0$ is the LOS path and the others are the N-LOS paths. The signal $y$ is transmitted by the RF front-end into the channel and received by the receiver. The output of the channel is the result of the convolution operation of the channel input and the channel response. This process can be expressed as
{
    \begin{equation}
      \tilde y_t = \sum_{i=0}^{n-1}{\left(y_{t-i}\cdot h_i\right)} + noise
    \end{equation}
}
where $y_t$ and $\tilde y_t$ are the input and output of the channel at time $t$, respectively. $noise$ is the additional white Gaussian noise. The receiver demodulates and decodes the wireless signal to recover the data stream $\hat u$. When recovering the data stream from the wireless signal, bit errors occur at the receiver due to the signal distortion caused by the multipath effect and the noise.

Unlike conventional trellis-based decoders, our model takes into account the interference of the wireless signal by the channel and noise, quantified by model parameters, as shown in \hyperref[fig1.a]{Fig. 1(b)}. We propose a hybrid HMM decoder. Specifically, the trellis-like structure of the convolutional code is modeled by the Markov process. However, due to the channel interference, the observations are not unique, so the Markov process can be modeled as an HMM whose parameters describe the channel prior. The discrete HMM model does not allow for soft inputs. To benefit from the soft decision, the Gaussian mixture model is introduced as the HMM observation to approximate the probability distribution of the demodulator likelihood output. GMM-HMM is a continuous variable based model that allows the decoder to accept soft inputs. Our method offers significant performance advantages over conventional decoding, regardless of whether the decision is hard or soft. In terms of the effects of nonlinear hardware effects, our method is also effective.

Our system is divided into two stages of training and testing:

Training:  In our system, the HMM is trained using supervised learning, which requires specific preparation of the dataset, but this can be done offline and does not affect the implementation. The specific details are presented in Section \ref{Training Hidden Markov Model by Supervised Learning}. After implementing K-Means clustering, the Gaussian mixture model is trained by the EM algorithm and converges to ideal parameters, Section \ref{Training GMM Model with Unsupervised Learning} shows details.

Testing: The decoding of the model adopts the Viterbi decoding algorithm\cite{forney1973viterbi} as the same as the conventional method, which does not increase computational cost. In hard decision mode, the decision result of the demodulator is converted into an observation sequence and used as the input of the decoder, which is presented in Section \ref{Decoding by the HMM Model}. In soft decision mode, the likelihood of the demodulator calculated by the cross-entropy is used as the soft input of the decoder, which is presented in Section \ref{Decoding by GMM-HMM}.

\section{HMM DECODER OF CONVOLUTIONAL CODES}
\label{DECODING CONVOLUTIONAL CODES WITH HMM}
In this section, we reconstruct the convolutional code structure using a Hidden Markov Model (HMM) and analyze why the HMM decoder is superior in multipath channels. We then describe the training process and the decoding algorithm for the model parameters.
\subsection{Reconstruct the Convolutional Code}
\label{Reconstruct the Convolutional Code}
\begin{figure*}[!t]
  \centering
  \includegraphics[width=6.25in]{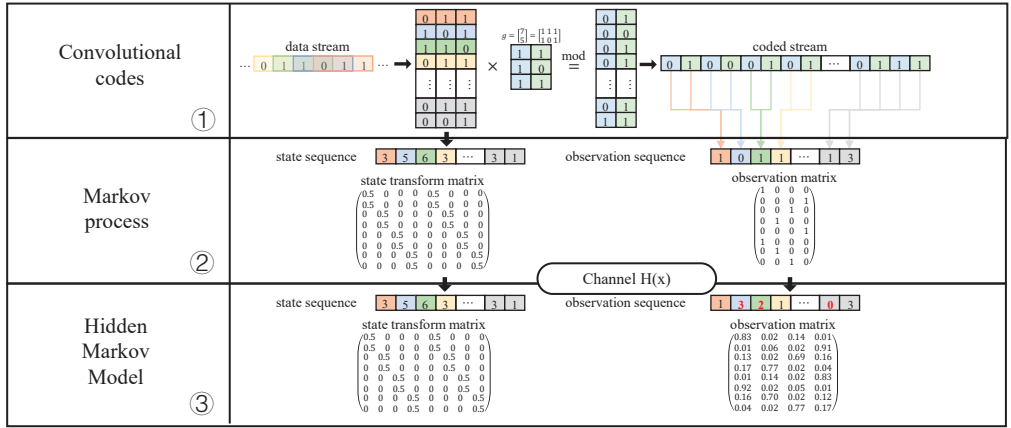}
  \caption{The process of $(2,1,3)$ convolutional code being reconstructed into HMM.}
  \label{fig3}
\end{figure*}

\hyperref[fig3]{Fig. 3} shows the reconstruction process of the $(2,1,3)$ convolutional code where $K=3$, $P=2$, and the generator polynomial functions are $g=[7\;5]$. The data stream is a random Markov chain since the state transition probability is 0.5. Consider a convolutional encoder with a constraint length of $K$ and a coding rate of $1/P$. The value of the $K$-stage shift register at time $t$ can be expressed as a binary vector $i_t=(x_1,x_2,\cdots,x_K)$, corresponding to the state of the HMM at time $t$. Since the $K$-stage shift register can generate $N=2^K$ possible values, the state set of the HMM is $Q=\{ q_1,q_2,\cdots,q_N \}$. The coded stream output by $P$ generator polynomial functions at time $t$ is represented as a binary vector $o_t=(y_1,y_2,\cdots,y_P)$ corresponding to the observation generated by the HMM state $i_t$ at time $t$. $P$ generator polynomial functions can generate $M=2^P$ different coded streams. Therefore, the observation set of the HMM is $V=\{ v_1,v_2,\cdots,v_M \}$. The data stream slips through the $K$-stage shift register, changing the value of the shift register, which represents the state transition of the HMM. The code stream generated by the generator polynomial function is expressed as an observation of the HMM.

Assuming that the channel is perfect and has no signal distortion, this does not cause bit errors in receiver demodulation. From the point of view of a state diagram, the state and the observation of the model are one-to-one. This is a Markov process, as shown in \hyperref[fig3]{Fig. 3}. However, the channel, whether AWGN or multipath, interferes with the signal and causes bit errors in the demodulator. Bit errors turn the one-to-one between states and observations into one-to-many, that is, multiple observations can be generated from one state. At this point, the state of the model has become a hidden variable, so the model is a Hidden Markov Model (HMM). Fortunately, we can use the observation matrix to describe the connection between the state and the observation.
Due to intersymbol interference in multipath channels, multiple states jointly determine an observation. This is not consistent with the assumption of observation independence of the HMM (assumption 2). However, symbols propagating on non-line-of-sight (N-LOS) paths (multipath channels) are attenuated much more than on line-of-sight (LOS) paths. Therefore, the main factor determining the observation is the direct path rather than the multipath channel, implying that the observation is mainly determined by the current state. Although assumption 2 is not matched, the HMM decoder can still be implemented on convolutional codes. A detailed explanation for this is illustrated in Section \ref{DISCUSSION AND LIMITATIONS}. The HMM decoder achieves a higher coding gain than the Viterbi decoder in multipath channels because the channel state information is parameterized.
In the next section, we will present how to train the parameters of the HMM.

\subsection{Training Hidden Markov Model by Supervised Learning}
\label{Training Hidden Markov Model by Supervised Learning}
The Hidden Markov Model has two training methods: unsupervised and supervised learning. Unsupervised learning is typically used in natural language processing. The well-known Baum-Welch algorithm is an expectation-maximum algorithm used in most training tasks for HMM models. 
However, the EM algorithm is completely random. In particular, different initial values lead to different results. To make matters worse, the bit error of the demodulator is used as a training data set, which leads to a deviation in the result. In our system, the supervised learning method is used to train the HMM model. The supervised learning method has less algorithm complexity than EM and can avoid randomness. However, it requires us to prepare the data set for training. Fortunately, the training process can be implemented offline, which does not hinder the application of our model. 

The training set $S = \{ (O_1, I_1),(O_2, I_2),\cdots,(O_S, I_S) \}$ contains the observation sequence and the corresponding state sequence, and the maximum likelihood estimation method can be applied to obtain the model parameters. The state transition matrix is estimated as:
\begin{equation}
  \hat a_{ij}=\frac{A_{ij}}{\sum_{j=1}^NA_{ij}},\quad i,j\in\{1,2,\cdots,N\}
\end{equation}
where $A_{ij}$ is the number of transitions from state $q_i$ to state $q_j$ in samples. The sum of each row of the state transition matrix must be equal to $1$ because the sum of the probabilities of transition from each state to other states is $1$. The observation matrix is estimated as:
\begin{equation}
  \hat b_{jk}=\frac{B_{jk}}{\sum_{k=1}^MB_{jk}},\quad j\in\{1,2,\cdots,N\},\quad k\in\{1,2,\cdots,M\}
\end{equation}
where $B_{jk}$ is the number of observations $v_k$ generated from the state $q_j$ in samples. The sum of each row of the observation matrix must equal $1$ because the probability of generating all observations from each state is $1$. The initial state vector $\hat \pi_i$ is the rate of state $q_i$ in $S$ samples.

\subsection{Decoding by the HMM Model}
\label{Decoding by the HMM Model}
The Viterbi algorithm for dynamic programming was preposed to decode graph models with hidden variables such as HMM \cite{forney1973viterbi}. Here we discuss how the Viterbi algorithm is used in the HMM decoder.

The state and observations of the HMM model are all generated by the data stream or the coded stream, which is a binary vector. However, the Viterbi algorithm requires that these elements be quantified in numbers to ensure that the algorithm executes normally. In our system, the binary vector is cleverly quantized into a decimal number for both the state and the observation of the HMM. As in the convolutional encoding process, the state of the HMM is generated by the data stream sliding through the $K$-stage shift register. Therefore, the $1st$-stage register records the entire data stream. After the state expressed as a decimal number is an output by the Viterbi decoding algorithm, we convert decimal to binary and select the first bit to obtain the decoded data stream.

The Viterbi algorithm uses the recursion method to calculate the probability for each path. Therefore, the accumulation of probabilities may lead to underflow if the data stream is too long.To tackle this problem, we apply Log-Viterbi, which requires logarithmic operation for each probability. We also normalize the probability for each path. With this improvement, the algorithm can be implemented stably even for long data streams.

\section{SOFT-DECISION BASED ON GMM-HMM}
\label{DECODING CONVOLUTIONAL CODES WITH GMMHMM}
In the previous section, we presented a novel HMM decoder and analyzed the advantages of this method for multipath channels. However, the conventional convolutional code provides a soft-decision decoding method that can be decoded according to the likelihood output of the demodulator. The soft-decision Viterbi algorithm achieves great performance gains because soft-input provides a likelihood to the decoder. However, the HMM decoder, a discrete model, cannot make soft decisions.

In this section, we use the Gaussian mixture model as the HMM observation. The elements in the observation set are converted from a binary vector to a Gaussian mixture probability distribution. GMM-HMM is a multidimensional model with continuous variables, so soft-decision decoding can be implemented.
\subsection{Convolutional Coding Based on GMM-HMM}
\label{Convolutional Coding Based on GMM-HMM}
The Gaussian mixture model is a parameterized model of a probability distribution that  aims to construct the probability distribution of an $n$-dimensional data set as a weighted superposition of $K$ Gaussian distributions. The probability density function is as follows:
\begin{equation}
  P(y|\theta)=\sum_{k=1}^K\alpha_k\phi(y|\theta_k)
\end{equation}
where $\alpha_k$ is the weight of the $k$-th Gaussian distribution and $\alpha_k\geq 0$, $\sum_{k=1}^K\alpha_k=1$. $\phi(y|\theta_k)$ is the probability density of the Gaussian distribution:
\begin{equation}
\phi(y|\theta_k)=\frac{1}{\sqrt{2\pi}\sigma_k}{\exp}(-\frac{(y-\mu_k)^2}{2\sigma_k^2})
\end{equation}
$\theta_k=(\mu_k,\sigma_k^2)$ is the $k$-th sub-model of the Gaussian mixture model.

In Section \ref{Reconstruct the Convolutional Code}, an HMM decoder is presented that reconfigures convolutional coding. Specifically, the coded stream is represented as a binary vector, which is viewed as an observation of the HMM model. However, when the wireless signal is transmitted to the receiver and demodulated, the demodulator cannot guarantee the correctness of the demodulation result. Probability is usually used to describe the degree of acceptability of the estimation result. Therefore, we approximate the probability distribution given by the demodulator with a Gaussian mixture model. Accordingly, the binary vector in the observation set of the HMM model is replaced by the GMM. The GMM-based hybrid HMM no longer relies on the demodulator's hard decision result, but predicts the state sequence using the Viterbi decoding algorithm based on the probability distribution. Note that the trained Gaussian mixture model contains the channel state information (CSI), so it can achieve better coding gains than SOVA. In the following, we will explain how the Gaussian mixture model is trained and what adjustments need to be made in the decoding phase.

\subsection{Training GMM Model with Unsupervised Learning}
\label{Training GMM Model with Unsupervised Learning}
The maximum likelihood estimation (MLE) approach is most commonly used to estimate the parameters of the Gaussian distribution. However, for the Gaussian mixture model, a weighted superposition of several Gaussian distributions, the parameters cannot be estimated directly using the MLE method, as this is not allowed in mathematics. Therefore, the EM algorithm based on the iterative estimation method is applied to the parameter estimation of the GMM model.

The essence of the EM algorithm is to update the parameters $\theta$ cyclically to maximize the expected value of the model. 
\begin{equation}
\theta^{(i+1)}={\mathop{\arg\max}_{\theta}}E_z[{\log}P(Y,Z|\theta)|Y,\theta^{(i)}]
\end{equation}
where $Y$ is the observed variable (i.e. observation), $Z$ is the hidden variable (i.e. state), and $P(Y,Z|\theta)$ is the joint probability distribution. Define $Q(\theta,\theta^{(i)})=E_z[{\log}P(Y,Z|\theta)Y,\theta^{(i)}]$ as the expectation of the model, and the algorithm essentially consists of two steps. In the $E$-step, the expectations of the model are calculated:
\begin{equation}
  \begin{split}
    Q(\theta,\theta^{(i)})&=E_z[{\log}P(Y,Z|\theta)|Y,\theta^{(i)}]\\
    &=\sum_{k=1}^K \bigg\{ n_k{\log}\alpha_k+\sum_{j=1}^N\hat \gamma_{jk}[{\log}(\frac{1}{\sqrt{2\pi}})\\&-{\log}\sigma_k-\frac{1}{2\sigma_k^2}(y_i-\mu_k)^2] \bigg\}
  \end{split}
\end{equation}
where $\gamma_{jk}$is defined as responsiveness:
\begin{equation}
  \gamma_{jk}=\frac{\alpha_k\phi(y_j|\theta_k)}{\sum_{k=1}^K\alpha_k\phi(y_i|\theta_k)}
\end{equation}

In the $M$-step, maximize Q and update model parameters:
\begin{equation}
  \theta^{(i+1)}=arg\max_{\theta}Q(\theta,\theta^{(i)})
\end{equation}
For a Gaussian mixture model, $\theta=(\alpha_k,\mu_k,\sigma_k^2)$ can be estimated as:
\begin{equation}
  \alpha_k=\frac{\sum_{j=1}^N\hat \gamma_{jk}}{N},\quad k=1,2,\cdots,K
\end{equation}
\begin{equation}
  \mu_k=\frac{\sum_{j=1}^N\hat \gamma_{jk}y_j}{\sum_{j=1}^N\hat \gamma_{jk}},\quad k=1,2,\cdots,K
\end{equation}
\begin{equation}
  \sigma_k^2=\frac{\sum_{j=1}^N\hat \gamma_{jk}(y_j-\mu_k)^2}{\sum_{j=1}^N\hat \gamma_{jk}},\quad k=1,2,\cdots,K
\end{equation}
Repeat this process until the model converges. 

Since the result of the EM algorithm is related to the initial value, it may cause the model not to converge if the initial value for training is chosen randomly. Therefore, the clustering result of the K-means algorithm is used as the initial parameter of the EM algorithm. \hyperref[alg1]{Algorithm 1} shows the training process of the GMM model.
\begin{algorithm}
	\caption{Training GMM by EM-Algorithm}
	\label{alg1}
	\begin{algorithmic}[1]
	  \STATE {Input:} 
      $$\begin{aligned} 
          & number\;of\;mixture:\;Q \\
          & training\;samples:\;O=(o_1,o_2,\cdots,o_T)
        \end{aligned}$$
		\STATE Initialization:
      $$\theta_k^0 \gets Kmeans(O,Q),\quad k=1,2,\cdots,K$$
        
		\STATE {While $\theta^{(i)}-\theta^{(i-1)}>\epsilon$ and $i\leq epochs$ Do:}
      $$\begin{aligned}
          & \hat \alpha_k \gets \frac{\sum_{j=1}^N\hat \gamma_{jk}}{N},\quad k=1,2,\cdots,K\\
          & \hat \mu_k \gets \frac{\sum_{j=1}^N\hat \gamma_{jk}y_j}{\sum_{j=1}^N\hat \gamma_{jk}},\quad k=1,2,\cdots,K\\
          & \hat \sigma_k^2 \gets \frac{\sum_{j=1}^N\hat \gamma_{jk}(y_j-\mu_k)^2}{\sum_{j=1}^N\hat \gamma_{jk}},\quad k=1,2,\cdots,K
      \end{aligned}$$
    \STATE {Output:} $\hat \theta=(\hat \alpha_k,\hat \mu_k,\hat \sigma_k^2),\quad k=1,2,\cdots,K$
	\end{algorithmic}  
\end{algorithm}

\subsection{Decoding by GMM-HMM}
\label{Decoding by GMM-HMM}
For the HMM model of discrete variables, we express the state and observations in terms of decimals. However, the probability output of the demodulator is a continuous variable $o_i=(\hat y_1,\hat y_2,\cdots,\hat y_P),\; y_i\in [0,1]$, which obviously cannot be quantized into a decimal number. Fortunately, GMM-HMM is a multidimensional model with continuous variables.  Therefore, the $P$-dimensional vector of the demodulator-likelihood only needs to be cross-entropy with each element in the observation set:
\begin{equation}
  \begin{aligned}
    o_i(t)&=CrossEntropy(o_i,\hat o(t))\\
    &=\prod_{n=1}^P\hat p_n^{y_n}\cdot(1-\hat p_n)^{(1-y_n)},\quad i=1,2,\cdots,M
  \end{aligned}
\end{equation}
where $o_i=\{0,1\}^P$ is the binary vector in the HMM observation set and $o(t)=(\hat p_1,\hat p_2,\cdots,\hat p_P)$ is the likelihood output of the demodulator at time $t$.

\section{EXPERIMENTAL RESULT}
\label{EXPERIMENTAL RESULT}
In this section, we evaluate the performance of the hybrid HMM decoder and compare the proposed method with the standard Viterbi decoder and neural network decoders in terms of bit error rate (BER) using numerical simulations.

\subsection{Simulation Settings}
\label{Simulation Settings}
We consider a multipath channel with additive white Gaussian noise (AWGN) containing one line-of-sight (LOS) path and two non-line-of-sight (N-LOS) paths. Assume the channel is a tapped delay line model used in \cite{ou2020neural,shlezinger2019viterbinet}. As mentioned in Section \ref{SYSTEM OVERVIEW}, the channel impulse response can be defined as:
\begin{equation}
  h=[h_0,h_1,h_2]=[1,e^{-\tau},e^{-2\tau}]
\end{equation}
where $\tau$ is the channel delay which is set to 0.7 in our experiment. We considered a variety of convolutional codes for generating functions in our experiments, and the specific details will be presented below. 

The experiment is performed based on the communication toolkit of MatLab 2021a. We adopt the BPSK modulation, and the demodulator uses the digital demodulation function of the communication toolbox. The HMM decoder was implemented based on the open-source Matlab-HMM toolbox \cite{matlab-hmm}. We evaluated our method on several $Eb/N0$. In each $Eb/N0$, we use 10,000 bits as training data and 1,000,000 bits as test data. The specific simulation settings are listed in \hyperref[tab1]{Table 1}.

\begin{table}[!t]
  \caption{Simulation Parameters}
  \centering
  \begin{tabular}{|c|c|}
  \hline
  Modulation Type & BPSK\\
  \hline
  Channel Impulse Response & $[1,e^{-\tau},e^{-2\tau}]$\\
  \hline
  Response Delay & 0.7\\
  \hline
  Training Dataset & 10,000 bits\\
  \hline
  Testing Dataset & 1,000,000 bits\\
  \hline
  Simulation framework & Communication Toolbox\\
  \hline
  Experiment Environment & MatLab 2021a\\
  \hline
  Experiment Equipment & Intel I5-10400F CPU\\
  \hline
  \end{tabular}
  \label{tab1}
\end{table}

\subsection{Performance of Proposed HMM decoder}
\label{Performance of Proposed HMM decoder}
\begin{figure}[!t]
  \centering
  \subfloat[]{\includegraphics[width=3.5in]{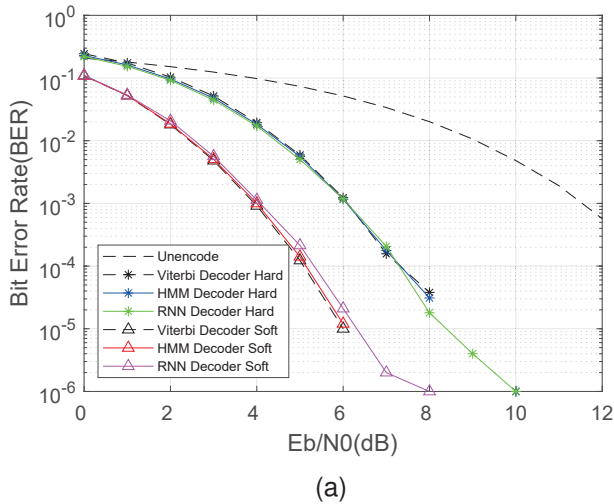}%
  \label{fig4.a}}
  \hfil
  \subfloat[]{\includegraphics[width=3.5in]{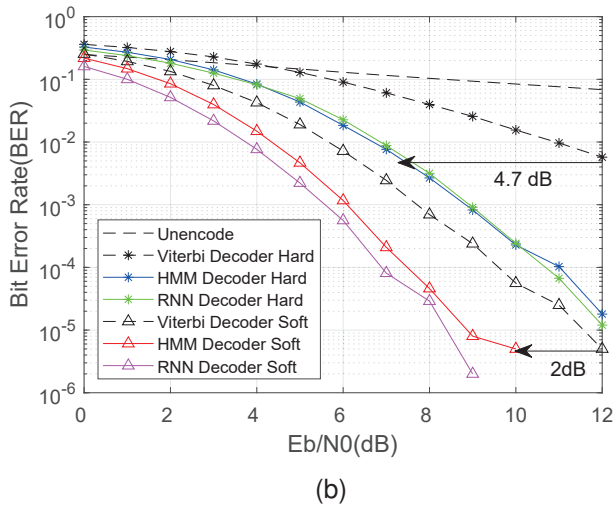}%
  \label{fig4.b}}
  \caption{The BER of the standard Viterbi Decoder, HMM decoder and RNN decoder\cite{kim2018communication} using hard-decision and soft-decision respectively in AWGN channel (a) and in multiple channel (b).}
  \label{fig4}
\end{figure}

In this experiment, we compare the BER of $(3,1,3)$ convolutional codes whose generator polynomial functions are $[7,7,5]$, respectively, using the standard Viterbi decoder and the HMM decoder under different $Eb/N0$. 

In the AWGN channel, there is no intersymbol interference (ISI) because the multipath effect does not exist. Conventional Viterbi decoders and HMM decoders can only derive the highest probability data stream based on the mathematical logic of the convolutional code. As a result, the two BER curves overlap as shown in \hyperref[fig4.a]{Fig. 4(a)}.

However, the ideal channel does not exist in the real environment. \hyperref[fig4.a]{Fig. 4(b)} compares the BER of the two decoders under a multipath channel. The HMM decoder performs significantly better than the Viterbi decoder under both hard- and soft-decision decoding. Specifically, there is a gain of 4.7 dB for the hard decision and a gain of 2 dB for the soft decision. The reason for this, as analyzed in Section \ref{Reconstruct the Convolutional Code}, is that the HMM decoder not only takes into account the mathematical structure of the convolutional code but also implicitly includes CSI in the model parameters.

\subsection{Comparison with RNN decoder}
\label{Comparison with RNN decoder}
Deep neural networks are used for channel decoding to improve the coding gain due to the nonlinear adaptability that arises from a large number of parameters. In this section, we compare the decoding performance, decoding latency, and model size of the RNN decoder\cite{kim2018communication} as well as HMM decoder. The evaluation results are obtained using the same test set.

\subsubsection{Decoding Performance}
\label{Decoding Performance}
To compare the performance of the decoders, we create the training sets for the RNN decoder and HMM decoder separately and maximize the performance of the two decoders. We compare the decoding performance of the RNN decoder and the HMM decoder in the AWGN channel and the multipath channel. As shown in \hyperref[fig4.a]{Fig. 4(a)}, the BER curves of both the RNN decoder and the HMM decoder overlap with the standard Viterbi decoder in the AWGN channel. This result is consistent with the result of \cite{kim2018communication}, indicating that the performance of the RNN decoder is maximized. 

The BER curve in the multipath channel is shown in \hyperref[fig4.a]{Fig. 4(b)}. In decoding with hard decisions, the coding gain of the RNN decoder and the HMM decoder is identical. With hard decisions, probability information is lost. Therefore, the decoding performance of the RNN decoder and the HMM decoder is limited due to hard input sequences. Compared to hard decisions, soft decision decoding receives more attention. The RNN decoder achieves almost 0.5dB higher coding gain than the HMM decoder. This is due to a large number of parameters and the complex nonlinear calculations in the RNN decoder. 

\subsubsection{Decoding Latency}

\label{Decoding Latency}
\begin{figure}[!t]
  \centering
  \subfloat[]{\includegraphics[width=1.8in]{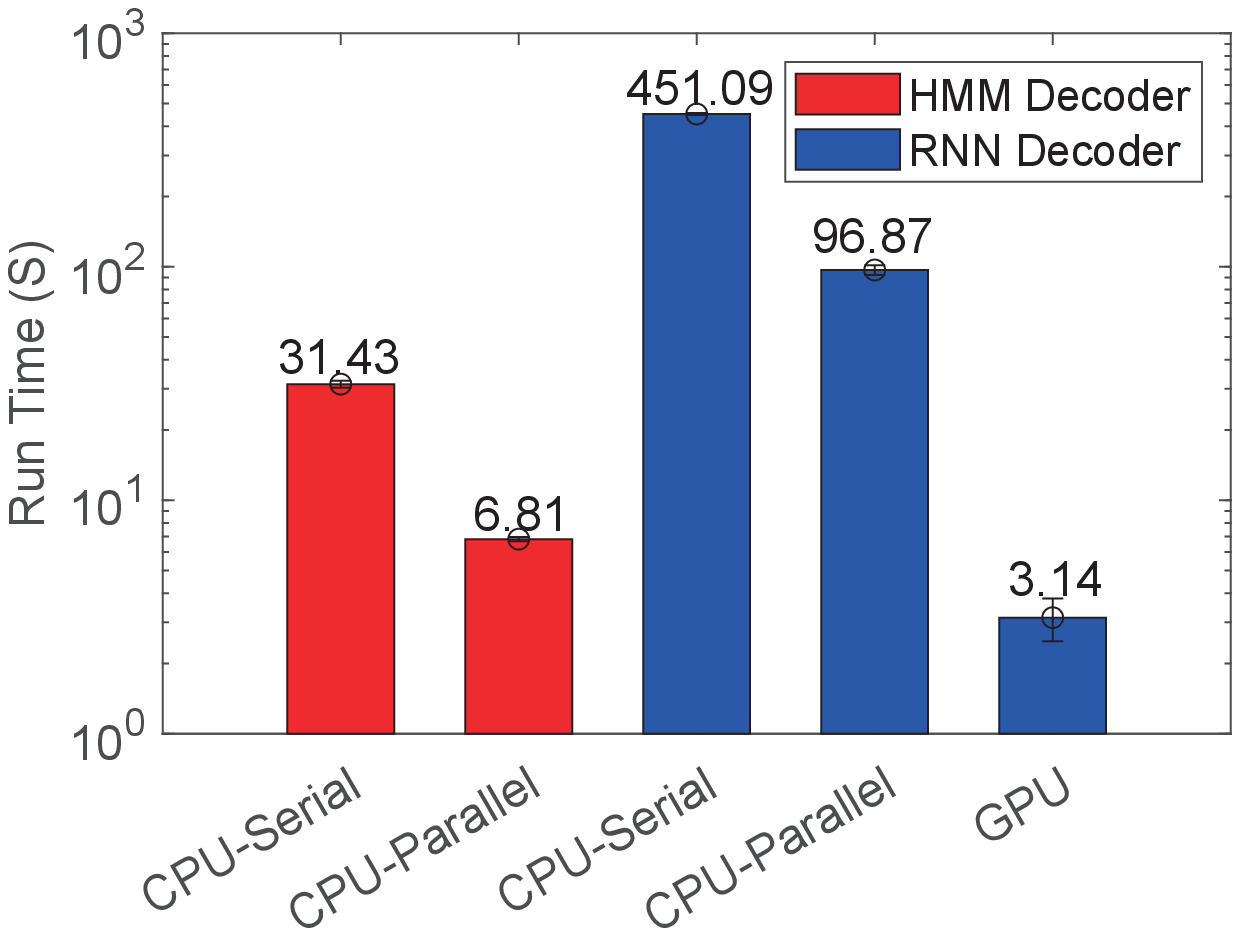}
  \label{fig5.a}}
  \subfloat[]{\includegraphics[width=1.8in]{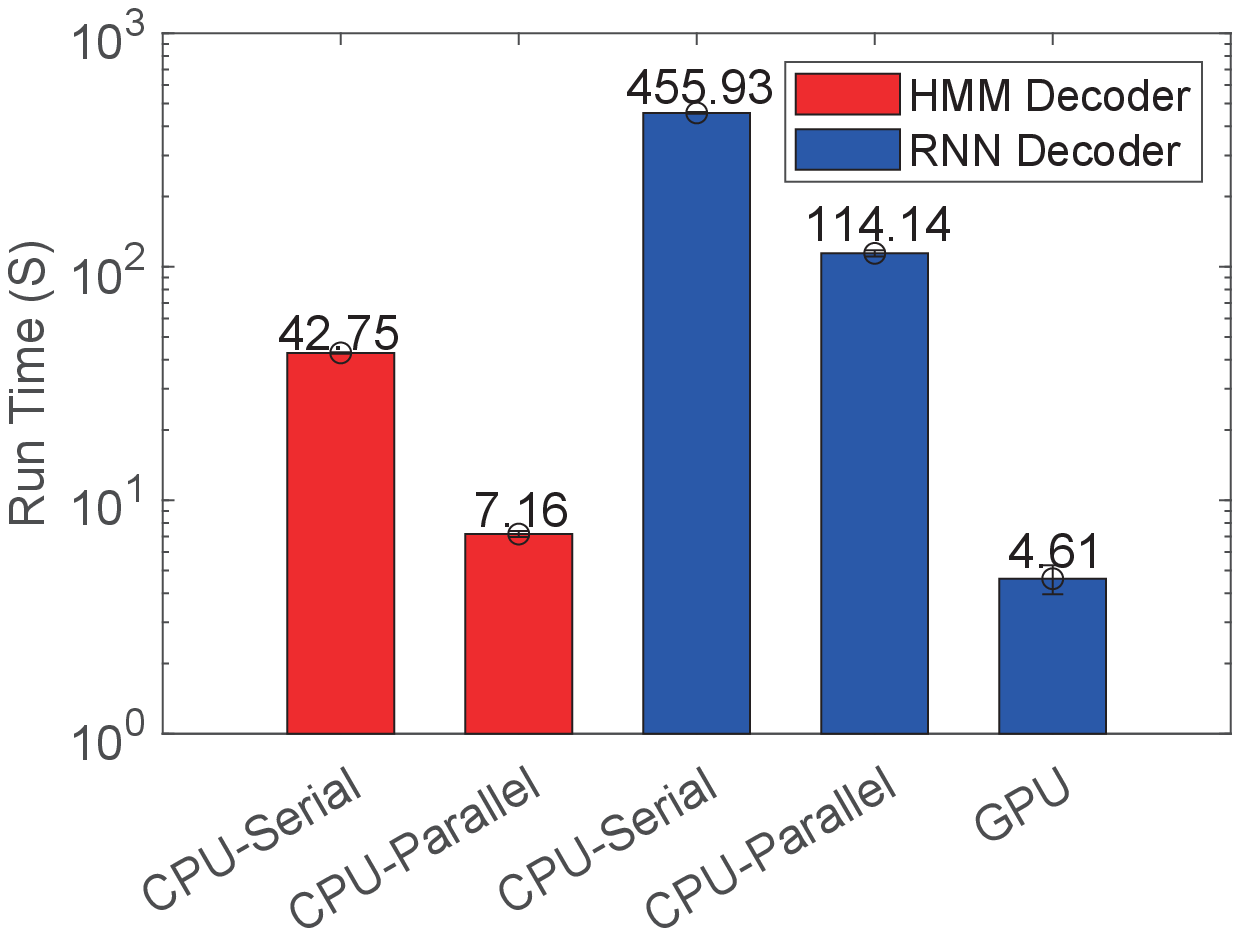}
  \label{fig5.b}}
  \caption{The Decoding latency of HMM decoder and RNN decoder using hard-decision (a) and soft-decision (b)}
  \label{fig5}
\end{figure}

\begin{table}[!t]
\centering
	\caption{Model size of HMM decoder and RNN decoder}
	\begin{tabular} {@{}ccc@{}}
		\toprule
		\textbf{Decoder} & \textbf{Total Params} & \textbf{Model Memory} \\ 
		\midrule
	       HMM Hard Decoder & 136 & 18 KB\\
	       
		   HMM Soft Decoder & 4744 & 57 KB\\
	       
	       RNN decoder & 3857601 & 14.716 MB\\
	       \midrule
	       HMM Hard Decoder for RSC & 136 & 18 KB\\
	       
		   HMM Soft Decoder for RSC & 4744 & 57 KB\\
	       
	       RNN decoder for RSC & 20526401 & 78.302 MB\\
     	 \bottomrule
     	 
	\end{tabular}\label{tab2}

\end{table}

We compare the latency of the RNN decoder and the HMM decoder when decoding 1,000,000 bits, as shown in \hyperref[fig5]{Fig. 5}. The numerical simulation results were obtained from the same PC. 
From the results, HMM decoder, regardless of whether it implements hard or soft decision decoding, saves $10\times$ more time than the RNN decoder when the CPU is configured serially or in parallel.
We also evaluate the latency of implementing the RNN decoder on a GPU (Nvidia Geforce RTX 2080Ti). The decoding latency is greatly reduced due to the parallel floating point processing capability of the GPU. However, this comes at the cost of additional cost and power consumption. 
Although the RNN decoder has an approximately 0.5 dB higher encoding gain than the HMM decoder, we suggest implementing the HMM decoder on edge computing due to lower computational requirements.

\subsubsection{Decoder Model Size}
\label{Decoder Model Size}
We compare the number of model parameters as well as the running memory consumption of the RNN decoder and the HMM decoder in \hyperref[tab2]{Table 2}. The model size of the HMM decoder is significantly smaller than that of the RNN decoder, suggesting that less running memory is required for the HMM decoder. Furthermore, this shows that the HMM decoder benefits from a smaller training set and a faster training time. The HMM decoder can be trained in just a few seconds, which is significantly lower than the ten minutes required by the RNN decoder on GPUs. Both the HMM decoder and the RNN decoder can be extended to Recursive Systematic Convolutional Codes (RSC). However, the RNN decoder requires a more complex neural network model than Convolutional Codes, while the HMM decoder requires no additional effort.

\subsection{Influence of Parameter of Convolutional Codes}
\label{Influence of Parameter of Convolutional Codes}
\begin{figure*}[!t]
  \centering
  \subfloat[]{\includegraphics[width=2.3in]{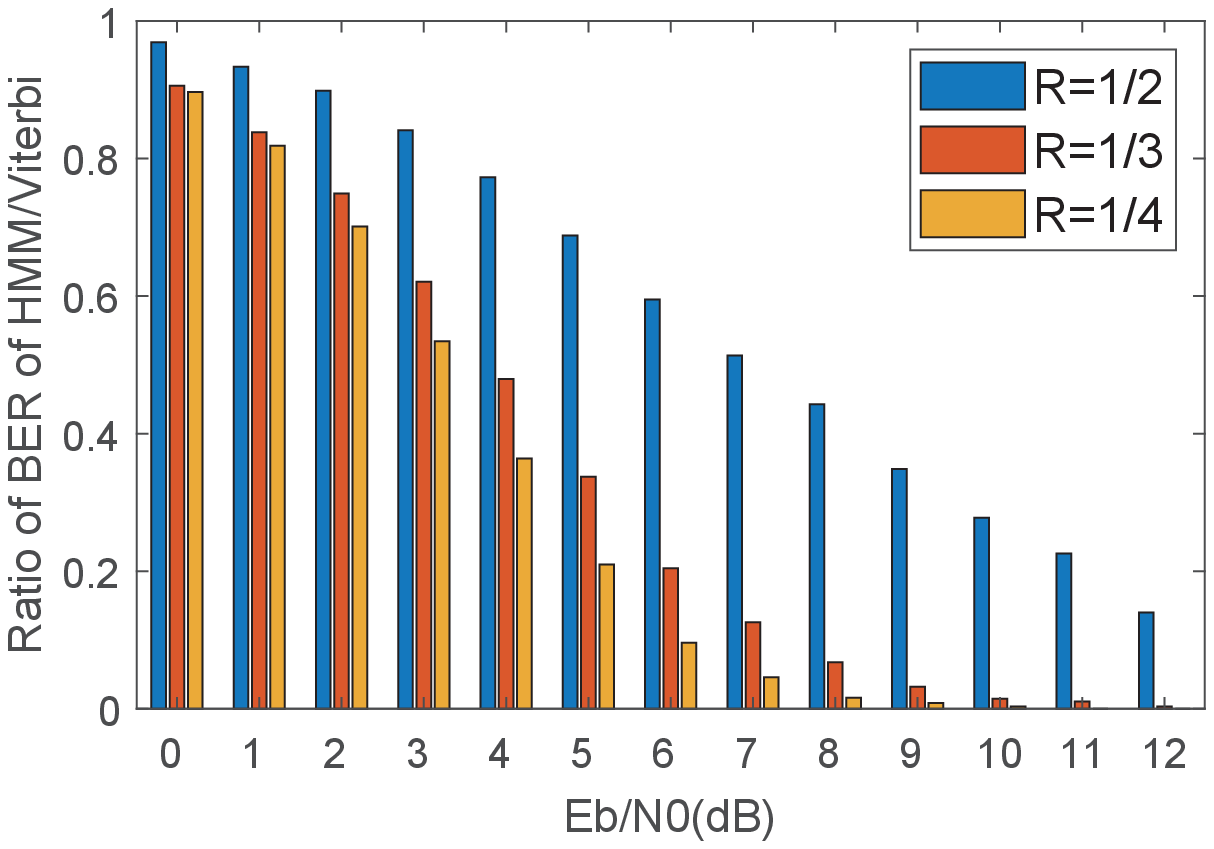}
  \label{fig6.a}}
  \subfloat[]{\includegraphics[width=2.3in]{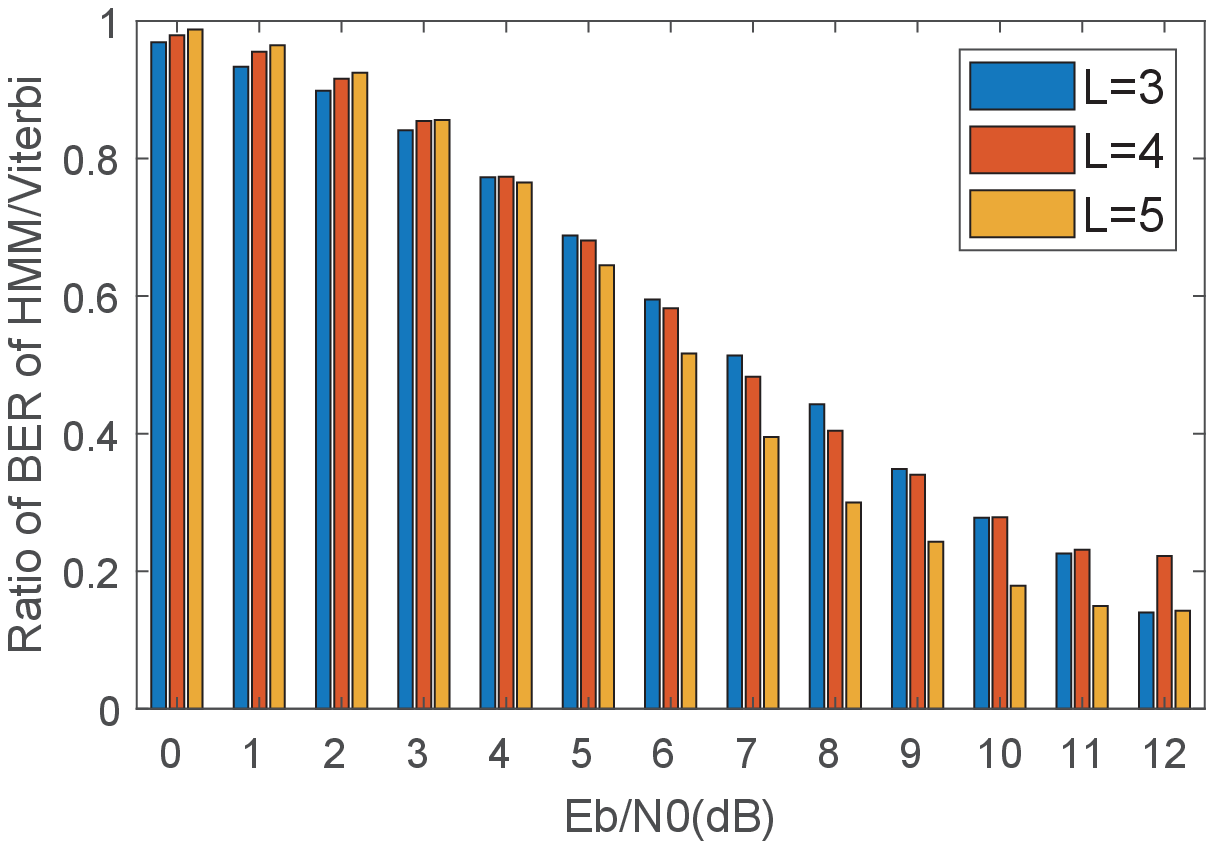}
  \label{fig6.b}}
  \subfloat[]{\includegraphics[width=2.3in]{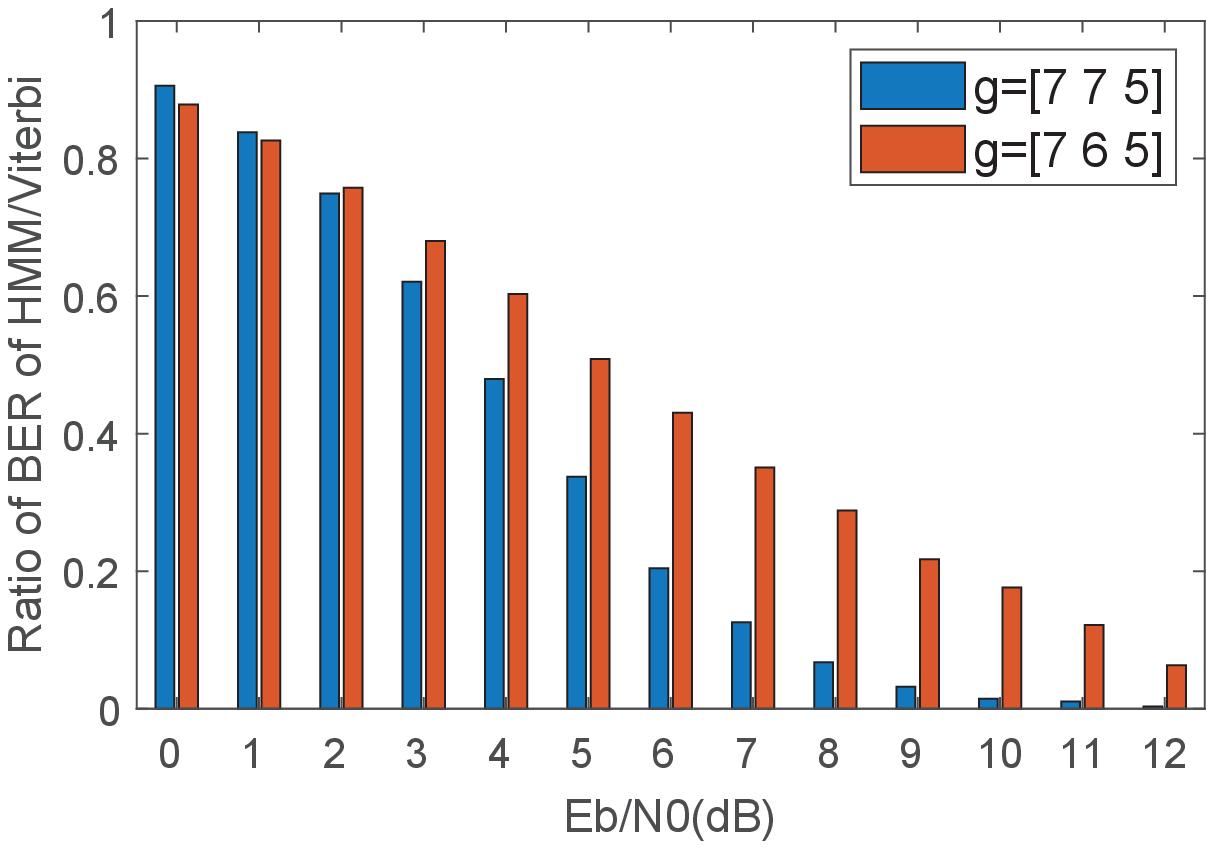}
  \label{fig6.c}}
  \hfil
  \subfloat[]{\includegraphics[width=2.3in]{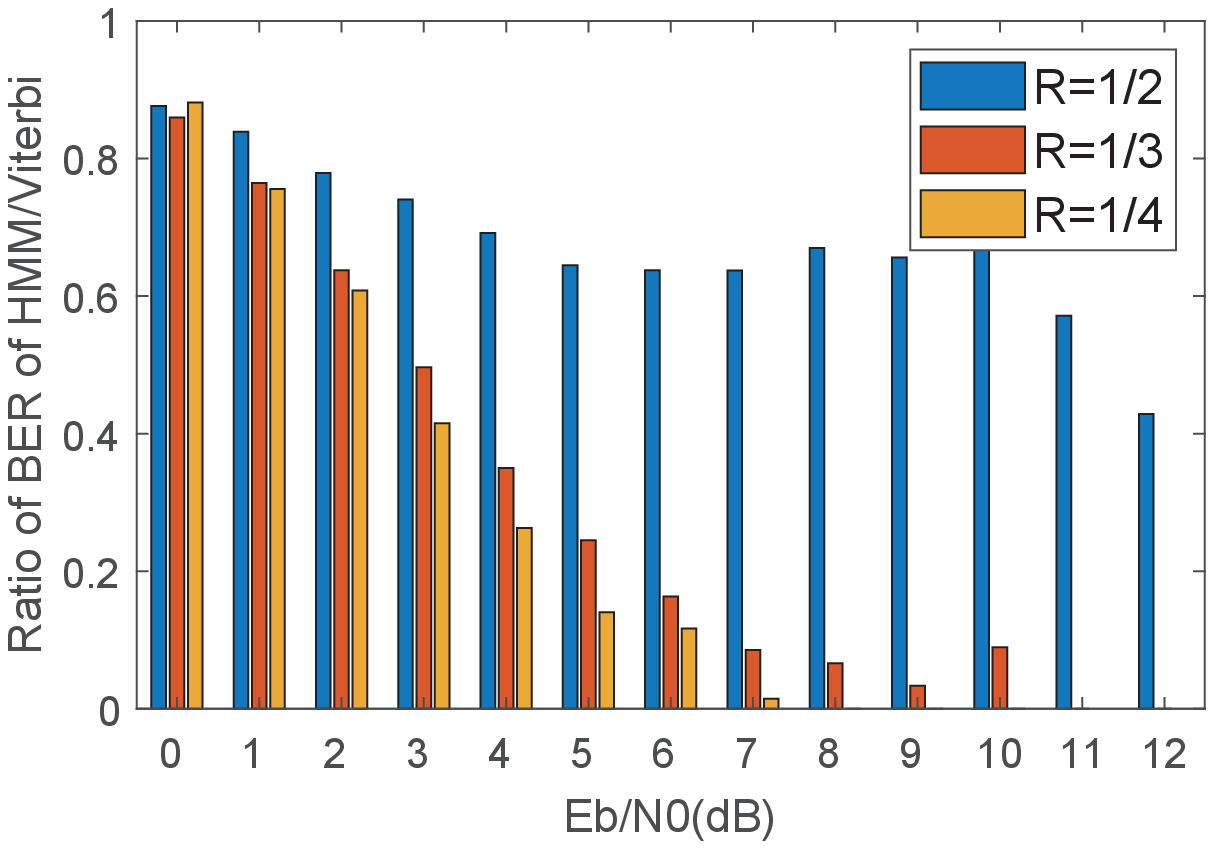}%
  \label{fig6.d}}
  \subfloat[]{\includegraphics[width=2.3in]{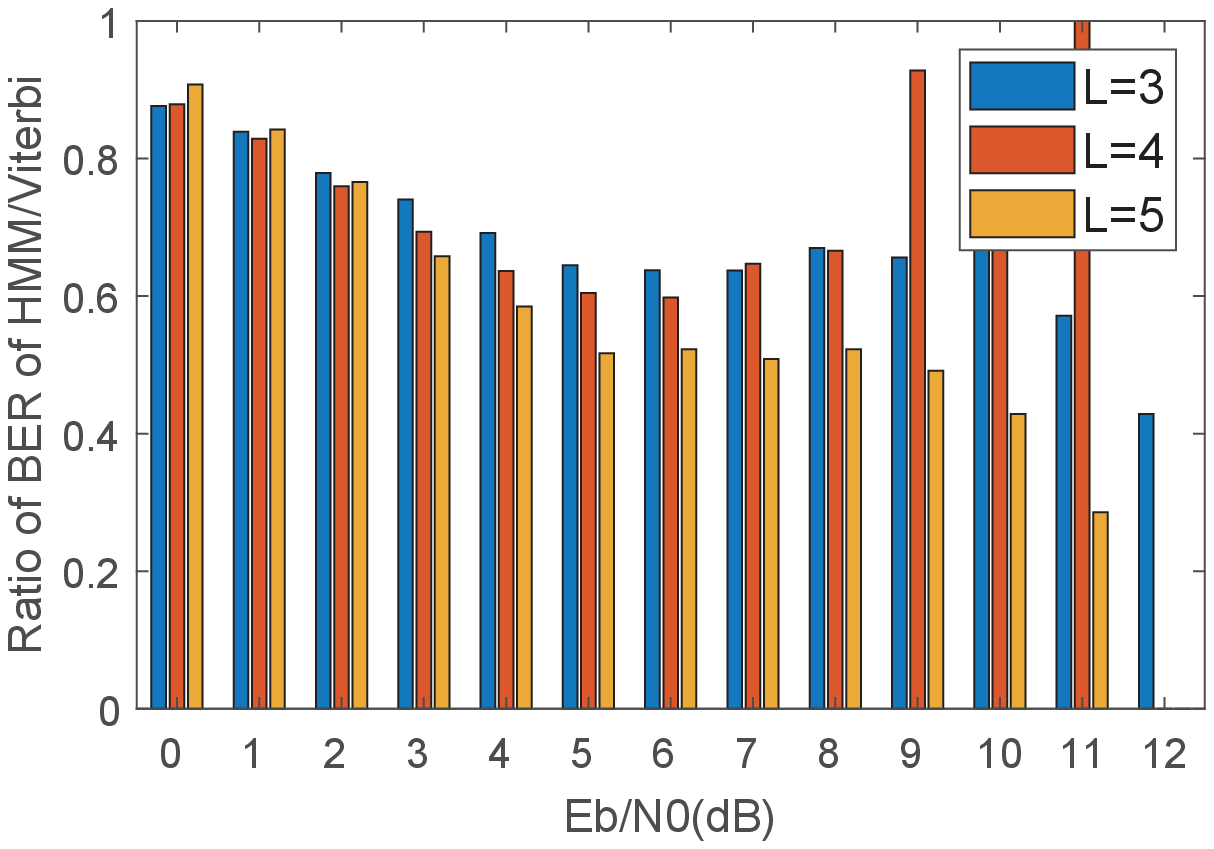}%
  \label{fig6.e}}
  \subfloat[]{\includegraphics[width=2.3in]{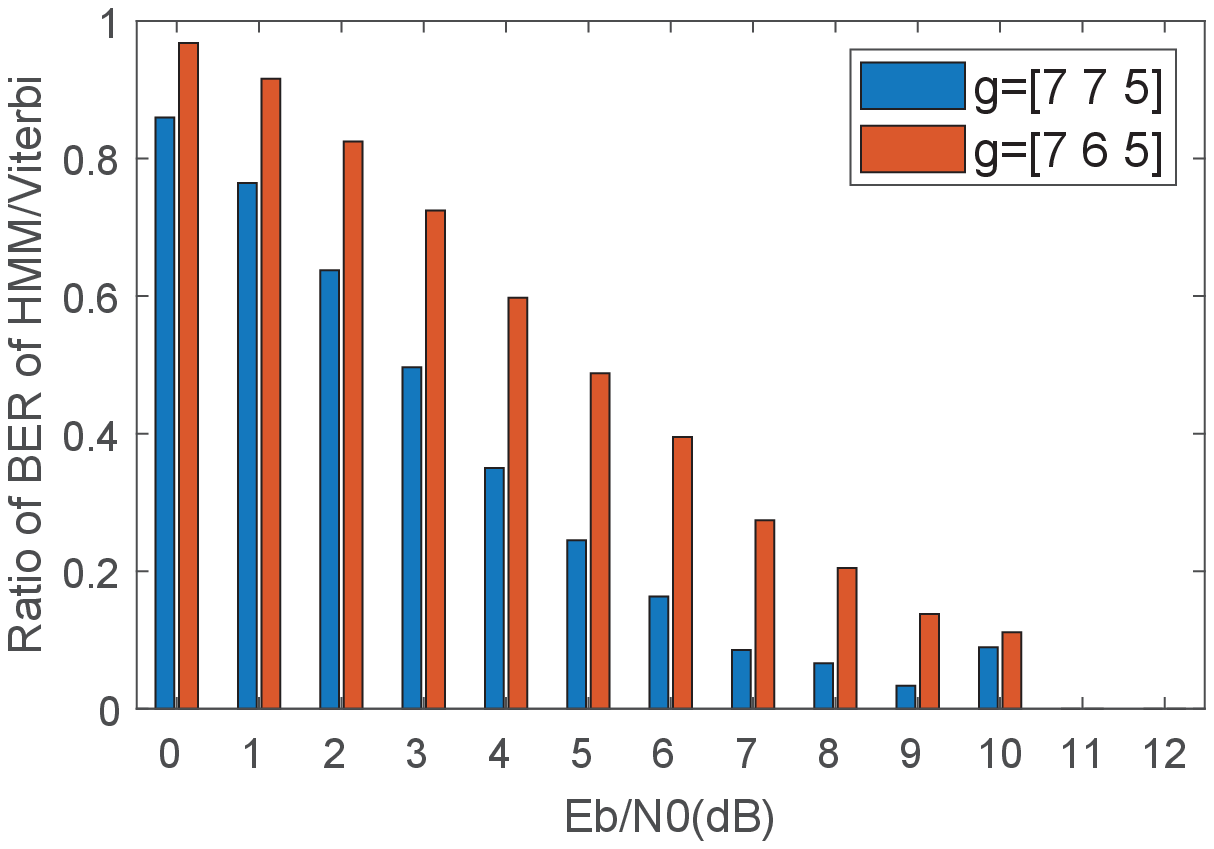}%
  \label{fig6.f}}
  \caption{Improvement of the hybrid HMM decoder in convolutional codes with different coding rates (a)(d), constraint lengths (b)(e), and generator polynomials (c)(f) respectively using hard-decision is shown in the upper row and soft-decision in the bottom row.}
  \label{fig6}
\end{figure*}
We evaluated the performance of the proposed model in convolutional codes with different coding rates, constraint lengths, and generator polynomials. The coding gains of convolutional codes with different parameters are inconsistent. We thus use the ratio of BER between the HMM decoder and the Viterbi decoder as the evaluation criterion, as shown in \hyperref[fig6]{Fig. 6}.
\subsubsection{Coding Rate}
\label{Coding Rate}
The coding rate of convolutional codes significantly affects the error correction performance of convolutional codes. Intuitively, convolutional codes with a low coding rate have large error correction potential because they have more redundancy. We investigated the performance improvement of the proposed model compared to the Viterbi decoder at different coding rates. Our experimental settings are based on convolutional codes with constraint length $K=3$, and three convolutional codes with a coding rate of $R=\frac{1}{2}$, $\frac{1}{3}$, and $\frac{1}{4}$ are evaluated, respectively, whose generator polynomials are $[7,5]$,$[7,7,5]$ and $[7,7,7,5]$. 

\hyperref[fig6.a]{Fig. 6(a)} and \hyperref[fig6.a]{6(d)} show the BER reduction rates for the hard decision and the soft decision using HMM decoder compared to the Viterbi decoder. We can conclude that the HMM decoder has a lower BER than the Viterbi decoder for different coding rates, regardless of whether the hard decision or the soft decision is used. In addition, the codes with a low coding rate can achieve better performance gains. A low coding rate means that the coding stream generated by a data bit is longer and there are more elements in the observation set. Low-rate convolutional codes have a complex mapping between the data stream and the coded stream, resulting in a lower BER. Complex codes give HMM decoders more opportunities for improvement. Therefore, the HMM decoder has better coding gains than complex convolutional codes.

\subsubsection{Constraint Length}
\label{Constraint Length}
The constraint length of the convolutional code is also an important factor affecting the error correction capability. We evaluate the convolutional codes with coding rate $R=\frac{1}{2}$ whose constraint lengths are $K=3,4,5$, as shown in \hyperref[fig6.a]{Fig. 6(b)} and \hyperref[fig6.a]{6(e)}.

The constraint length of the convolutional code is determined by the number of registers. Intuitively, increasing the number of registers means that the coding structure becomes more complex. As a result, the performance of the HMM decoder is improved more for convolutional codes with longer constraint lengths, as analyzed in the previous section.

\subsubsection{Generator Polynomial}
\label{Generator Polynomial}
We evaluated a convolutional code consisting of a variety of generator polynomial functions. 

As shown in \hyperref[fig6.a]{Fig. 6(c)} and \hyperref[fig6.a]{6(f)}, the HMM decoder has great improvements for convolutional codes with different generating polynomial functions.

\subsection{The Influence of Non-Linear Effects on Decoding}
\label{The Influence of Non-Linear Effects on Decoding}
\begin{figure}[!t]
  \centering
  \includegraphics[width=3.5in]{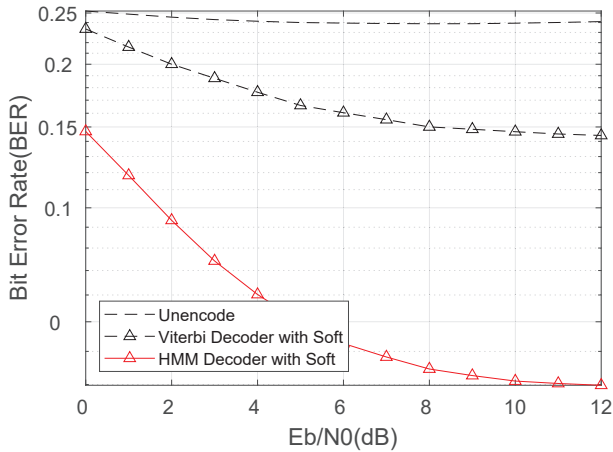}
  \caption{BER of HMM decoder and Viterbi decoder using soft-decision in multipath channel considering the non-linear effects of hardware.}
  \label{fig7}
\end{figure}
In the existing communication systems, the nonlinear effect caused by the hardware cannot be ignored. The disturbance of the signal by the hardware can also be considered as part of the channel response. Therefore, we evaluated the robustness of the method to this effect. We considered the following function to simulate the nonlinear effect of hardware devices, which is identical to \cite{xu2018joint,ye2017initial,olmos2009joint,salamanca2010channel}.
\begin{equation}
  g(v)=v+0.2v^2-0.1v^3+0.5\cos(\pi v)
\end{equation}

Due to the interference caused by the nonlinear effects of the hardware, the decision error of the demodulator increases sharply, which causes the performance of the decoder with hard decision decoding to drop sharply or even worse than without channel coding. Therefore, in this experiment, we only compare the performance of the HMM decoder and the Viterbi decoder with soft-decision decoding, as shown in \hyperref[fig7]{Fig. 7}. The HMM decoder provides a significant performance improvement over the Viterbi decoder because the nonlinear effects can be parameterized by the HMM decoder as part of the channel response.

\subsection{Expansion in Recursive Systematic Convolutional Codes}
\label{Expansion in Recursive Systematic Convolutional Codes}
\begin{figure}[!t]
  \centering
  \includegraphics[width=3.5in]{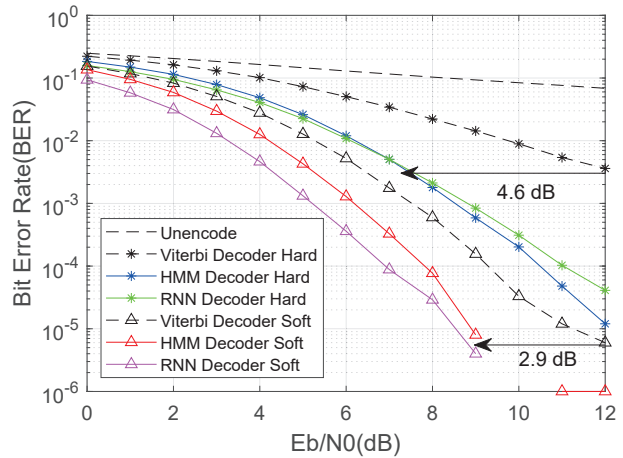}
  \caption{The validity of the hybrid HMM decoder on the RSC code.}
  \label{fig8}
\end{figure}
Recursive systematic convolutional codes are often implemented in cascaded coding systems such as turbo codes, which can further improve the error correction performance of convolutional codes and even approach the Shannon limit. We have verified the effect of the HMM decoder on RSC codes, as shown in \hyperref[fig7]{Fig. 8}. Our method also works for RSC codes. Therefore, the HMM decoder proposed in this paper can be used in Turbo decoders to further improve performance. The RNN decoder uses a bidirectional LSTM and more neurons in hidden layers to cope with the increased state dimension in RSC. However, the results are similar to \ref{Decoding Performance}, and the coding gain is about 0.5dB higher than HMM decoder, which does not require any additional extension.

\section{DISCUSSION AND LIMITATIONS}
\label{DISCUSSION AND LIMITATIONS}

This paper proposes an HMM decoder for convolutional codes that parameterizes channel priorities by pre-training and achieves higher coding gains than standard Viterbi decoders in multipath channels. Moreover, compared with neural network decoders, it has lower decoding latency and a smaller model size.  However, there are still areas where HMM decoders can be improved:

As discussed in Section \ref{Reconstruct the Convolutional Code}, multipath channels lead to intersymbol interference, breaking the assumption of observational independence (Assumption 2). That is, the observations are determined by several states. The HMM decoder tolerates this problem. Indeed, the observation independence assumption of HMM is a strong assumption that cannot be met in most practical applications. 
In natural language processing (NLP), for example, complex human language is always influenced by the meaning of the context. This does not mean that the HMM is ineffective. On the contrary, it is widely used in NLP.

The Maximum Entropy Markov Model (MEMM) and the Conditional Random Field (CRF) are proposed to improve the HMM. The MEMM abandons the assumption of observation independence, according to which observations are influenced by both the current state and the previous state. The CRF model further removes the assumption of homogeneity of state. 
We believe that MEMM and CRF can further improve the coding gain and even achieve the performance of neural network decoders. Although CRF implies a larger parameter scale than HMM, we believe that memory consumption can be significantly reduced compared to neural network decoders. We leave this interesting yet challenging topic for future work.

\section{CONCLUSION AND FUTURE WORK}
\label{CONCLUSION AND FUTURE WORK}
Convolutional codes are an effective way to solve the wireless communication reliability problem in edge computing. In this paper, we propose an HMM decoder that reconstructs the convolutional code and parameterizes the channel state information in the model.
To benefit from soft decision decoding, the GMM is introduced into the HMM decoder. We evaluated the effectiveness of the method through numerical simulations. Decoding with hard and soft decisions achieves 4.7 dB and 2 dB gain compared to the standard Viterbi decoder.
We evaluated the HMM decoder on convolutional codes with different parameters and concluded that the HMM decoder provides a more significant improvement for complex codes.
Hardware non-linear effects can also be parameterized to provide advantages over standard Viterbi decoder. We have demonstrated that the HMM decoder is effective on RSC codes, so it is foreseeable that the method can improve the performance of turbo codes. This is a topic for future research.

\section*{Acknowledgments}

We sincerely thank all reviewers for their insightful suggestions. This work was supported by the NSFC (National Natural Science Foundation of China) Youth Science Fund under Grant No. 61802309, the International Science and Technology Cooperation Projects of Shaanxi Province under Grand No. 2019KWZ-05.

\bibliography{reference}
\bibliographystyle{IEEEtran}

\vfill

\end{document}